\documentclass[12pt]{iopart}

\usepackage{booktabs}
\usepackage{xcolor}
\usepackage{bbold}
\usepackage{mathrsfs}
\usepackage[numbers, sort&compress]{natbib}
\usepackage{here}
\usepackage{graphicx}
\usepackage{caption,subcaption}
\newcommand{\red}[1]{{\color{red} #1}}

\usepackage{hyperref}
\hypersetup{plainpages=false,colorlinks=true,linkcolor=blue, citecolor=blue, urlcolor=blue}

\begin{document}

\title[Reflectance spectral studies of spark plasma sintered tungsten carbide pellet]{{Reflectance spectral studies of spark plasma sintered tungsten carbide pellet}}

\author{Toshiharu Chono$^{1}$, Hisashi Tokutomi$^{2}$, Kazuma Nakamura$^{1,3}$\thanks{E-mail: kazuma@mns.kyutech.ac.jp} \& Koji Miyazaki$^{1,3,4}$\thanks{E-mail: miyazaki.koji.962@m.kyushu-u.ac.jp}}

\address{$^{1}$Graduate School of Engineering, Kyushu Institute of Technology, 1-1 Sensui-cho, Tobata-ku, Kitakyushu, Fukuoka 804-8550, Japan \\
$^{2}$KOJUNDO CHEMICAL LABORATORY CO., LTD., 5-1-28, Chiyoda, Sakado, Saitama 350-0284, Japan \\
$^{3}$Integrated Research Center for Energy and Environment Advanced Technology, Kyushu Institute of Technology, 1-1 Sensui-cho, Tobata-ku, Kitakyushu, Fukuoka 804-8550, Japan \\
$^{4}$Graduate School of Engineering, Kyushu University, 744 Motooka, Nishi-ku, Fukuoka 819-0395, Fukuoka, Japan}
\ead{kazuma@mns.kyutech.ac.jp, miyazaki.koji.962@m.kyushu-u.ac.jp}
\vspace{10pt}
\begin{indented}
%\item[]August 2024
 \item[] June 2024
\end{indented}

\begin{abstract}
We report the first spectral reflectance of tungsten carbide (WC) as potential solar selective absorber. We developed an optical measurement system for visible to mid-infrared spectroscopy, covering the range of 0.1 to 2.5 eV, to evaluate the solar selectivity. A polycrystalline WC was prepared using spark plasma sintering method. The measured spectral reflectance of WC exhibits a low-energy plasma excitation around 0.6 eV corresponding to the cutoff energy of sunlight, consistent with {\em ab initio} calculations, thus making it preferable for the solar selective absorber. We also discuss effects of the sample quality on the spectral reflectance.
\end{abstract}
%
% Uncomment for keywords
%\vspace{2pc}
%\noindent{\it Keywords}: XXXXXX, YYYYYYYY, ZZZZZZZZZ
%
\vspace{-1cm} 
% Uncomment for Submitted to journal title message
\submitto{\JJAP}
\vspace{0.2cm} 
%
% Uncomment if a separate title page is required
%\maketitle
% 
% For two-column output uncomment the next line and choose [10pt] rather than [12pt] in the \documentclass declaration
%\ioptwocol
%
%\section{Introduction: file preparation and submission}
%\section{Introduction}\label{sec_Introduction}

Utilization of solar thermal systems is one of the key technologies that contributes to carbon neutrality through an effective use of solar energy~\cite{Shi_2022, kohiyama_2016}. Selective absorption of sunlight is crucial characteristic for maintaining high temperatures of the absorber heated by sunlight, thereby preventing thermal radiation loss \cite{Cao_2014}. Devices aimed at achieving high performance in this regard are actively under development. The energy range for radiative heat transfer typically is considered from 0.05 eV to 4.0 eV, making a spectrometer for quantitative measurements of optical spectra within this range extremely important.  Recent {\em ab initio} analysis for TiCN-based cermets, commonly used as cutting tools, has revealed {that a tungsten carbide (WC) being a primary component of the cermets exhibits a low-energy plasma excitation around 0.6 eV corresponding to the cutoff energy of sunlight~\cite{Hayakawa2023}.} While WC-based superalloys are widely used in various applications \cite{recycle_WC_2018, cemented_carbide_review_2019}, to the best of our knowledge, the spectral reflectance has not been measured, and understanding of their optical properties is still limited \cite{Lundstrom_1984,He_2017,Ma_2018,Hou_2019}. {There is a report on an optical measurement of WC using FT-IR~\cite{FTIR_study_of_WC}, but the energy range studied  -- 0.06 to 0.2 eV -- is insufficient to detect the 0.6-eV plasma edge of WC. Experimental verification of the {\em ab initio} result is highly desirable; the {\em ab initio} calculation is predictive, but it yields results under idealized conditions, such as the absence of impurities and lattice defects. Thus, it is nontrivial that the prediction matches a measured result for real WC. The agreement between theory and experiment is a firm starting point for materials design.}

{In this study, we present experimental results on spectral reflectance of a sinterd WC pellet. We developed an optical measurement system ranging from visible to mid-infrared regions, covering 0.1 to 2.5 eV. %With the energy range of sunlight from 0.6 to 4.0 eV and the thermal radiation range of materials from 0.05 to 0.6 eV, 
%\red{In research into solar absorption materials, it is necessary to consider a wide wavelength range from visible light to near infrared light, quantitative measurement within this range is crucial. Additionally, it is necessary to be able to properly characterize the plasma edge of the material.}} 
{In general, the spectral range of sunlight is 0.6 to 4.0 eV, and the radiation range of materials is 0.12 to 0.6 eV. Then, quantitative measurement within these range is crucial in discussing the solar selective  absorption performance of materials.} While many spectrometers measure only the infrared or visible regions separately, it is desirable to measure the entire range with a single device from a quantitative perspective \cite{makino_2001}.

A polycrystalline WC pellet was prepared by sintering powder samples. Typically, sintering powders with high melting points results in low material density, although such low density materials with rough surfaces can significantly degrade spectral reflectance by scattering. Hence, we prepared high density WC without pores using a spark plasma sintering (SPS) method \cite{SPS_tokita1999, haibin-SPS, jiang-SPS, Roberta-SPS}. The resulting spectral reflectance of WC exhibits a clear low-energy plasma edge around 0.6 eV (2.0 $\mu$m) just corresponding to the cutoff energy wavelength of sunlight, thus proposing WC as a promising solar selective absorber. We found that the experimental spectrum is in good quantitative agreement with {\em ab initio} calculations~\cite{Hayakawa2023}. Additionally, we discuss effects of sample quality on spectral reflectance by comparing samples prepared using different sintering methods. %Spectral radiation properties are dependent on both material properties and surface microstructures.

%\section{Method}\label{sec_method} 
%\subsection{Optical measurement system}\label{Optical measurement system}
%In the present study, we developed a spectrometer in the visible and near-infrared regions to evaluate the optical properties of solar selective absorber. 
%The energy range of sunlight is from 0.6 eV to 4.0 eV, and the thermal radiation from materials is ranging from 0.05 eV to 0.6 eV. Quantitative measurement of this energy range is important in the search for solar absorbing materials. There are many spectrometers that measure only the \tr{mid}-infrared region or only the visible region, but it is desirable to be able to measure this region with one device in terms of a quantitative point of view.

Figure~\ref{equipment}(a) shows our developed optical measurement system: The equipment mainly consists of a light source (visible and infrared), an optical chopper, a parabolic mirror, and a spectrophotometer. The light emitted from the light source is converted into intermittent light by an optical chopper and reaches the sample placed at the focal point of the parabolic mirror. The light reflected by the sample is converted into parallel light via the parabolic mirror and reaches a focal point via the reflection at the opposite-side parabolic mirror. The light reflected by a convex mirror at the focal point is thus transported to the spectrophotometer. %Our spectrophotometer consists of a filter, a diffraction grating, and a sensor. 
The light incident on the spectrophotometer is separated by a filter and a diffraction grating, and converted into voltage by a sensor, allowing the reflection intensity of the sample to be measured. In the measurements, noise is reduced by using an optical chopper and lock-in amplifier.
\begin{figure}[htpb] 
\begin{center} 
\includegraphics[width=0.55\textwidth]{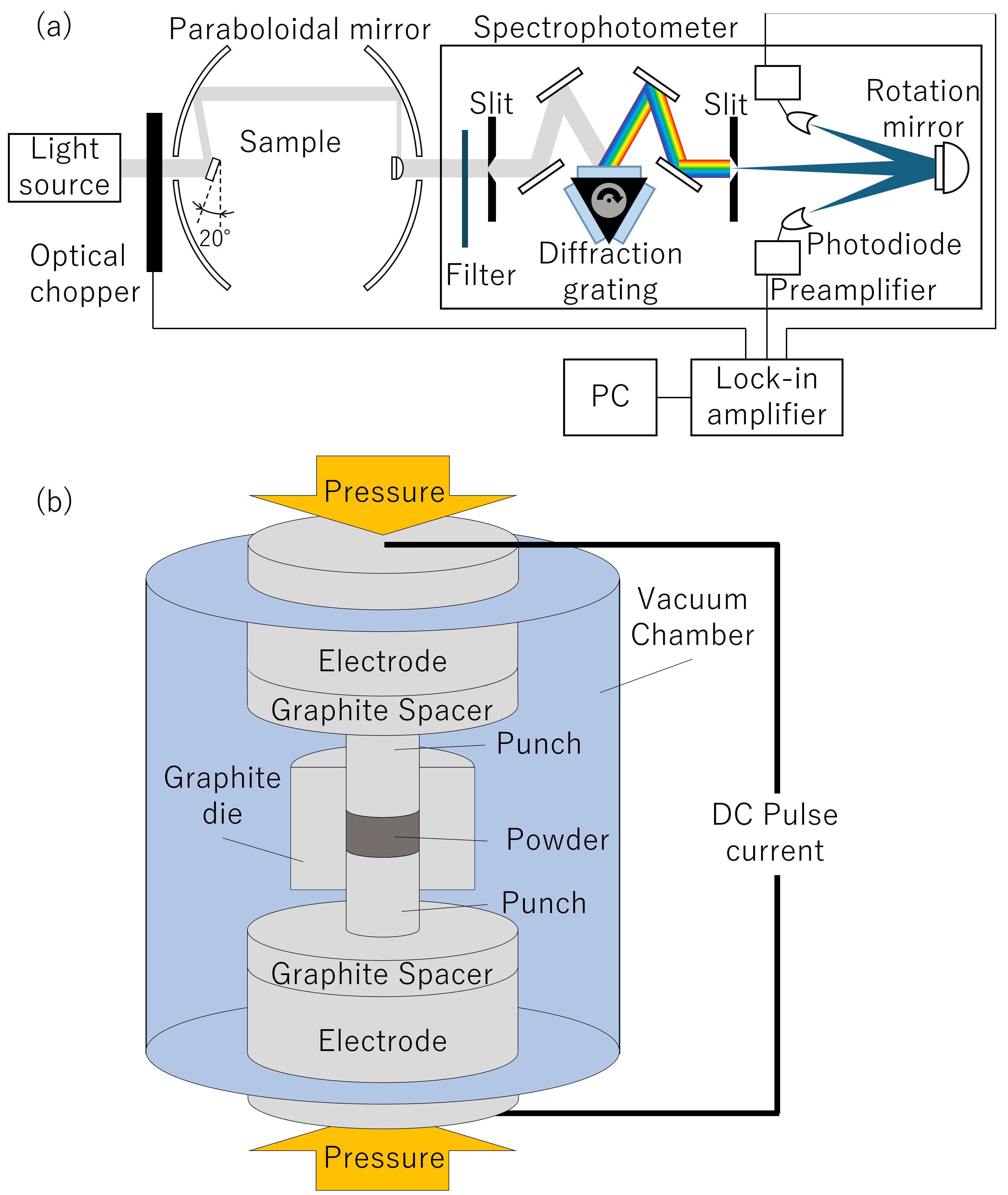}
%\includegraphics[width=0.75\textwidth]{Fig-1_equipment.jpg}
%\includegraphics[width=1\textwidth]{schematic.eps}%width=0.48
%\vspace{-0.5cm} 
\end{center} 
\caption{{Schematic diagrams of (a) our developed optical measurement system and (b) spark plasma sintering device to synthesize high-density sintered body.}}
%(a) A schematic diagram of our developed optical measurement system: The equipment mainly consists of a light source (visible and infrared), an optical chopper, a parabolic mirror, and a spectrophotometer. The light emitted from the light source is converted into intermittent light by an optical chopper and reaches the sample placed at the focal point of the parabolic mirror. The light reflected by the sample is converted into parallel light via the parabolic mirror and reaches a focal point via the reflection at the opposite-side parabolic mirror. The light reflected by a convex mirror at the focal point is thus transported to the spectrophotometer. Our spectrophotometer consists of a filter, a diffraction grating, and a sensor. The light incident on the spectrophotometer is separated by a filter and a diffraction grating, and converted into voltage by a sensor, allowing the reflection intensity of the sample to be measured. In the measurements, noise is reduced by using an optical chopper and lock-in amplifier. (b) A schematic diagram of spark plasma sintering (SPS) device mainly consisting of a chamber, electrodes, graphite spacers, punches, and graphite dies. By applying pressure and pulsed current heating to the powder samples, a bulk sintered body with high relative density can be obtained.}
\label{equipment}
\end{figure}

%The reflectance spectrum of the measured sample $R_{sample}$ is corrected using a reflectance spectrum of a calibration sample $R_{ref}$ as follows: 
%\begin{eqnarray}
% R_{sample}(\lambda) = \frac{I_{sample}(\lambda)}{I_{ref}(\lambda)} R_{ref}(\lambda)
%\label{eq:reflectance} 
%\end{eqnarray}
%with $I_{sample}$ and $I_{ref}$ being reflection intensities of the measured and calibration samples, respectively. 

%\subsection{Material synthesis by spark plasma sintering method}\label{spark plasma sintering}
Sintering powder samples of materials with high melting points is generally difficult. The resulting sintered body contains many pores, and such pores are cause of multiple scattering of light, which significantly degrades the quantitative accuracy of the reflection spectrum. Although it is desirable to remove the pores by heating or pressurization [hot press (HP) method], there are limitations for materials with the high melting points. In the present study, we use an SPS method to synthesize samples. In the SPS, since the powder sample is sintered while passing an electric current through it, there is a self-heating effect because of the Joule heat, which is effective in reducing the pores to a certain %\red{
amount.
%} %In the present study, we compare samples obtained by the conventional HP method with those obtained by the SPS method, and study how much the difference in the sintering method affects the reflectance spectra.

Figure~\ref{equipment}(b) is a schematic diagram of the SPS device (Fuji Electronic Industrial Co., Ltd.) mainly consisting of a chamber, electrodes, graphite spacers, punches, and graphite dies. By applying pressure and pulsed current heating to the powder samples, a bulk sintered body with the high density can be obtained.
%\begin{figure}[h!] 
%\begin{center} 
%\includegraphics[width=0.55\textwidth]{Fig-2_SPS_equipment.jpg}
%\includegraphics[width=0.4\textwidth]{Fig-2_SPS_equipment.jpg}
%\end{center} 
%\vspace{-0.5cm} 
%\caption{A schematic diagram of spark plasma sintering (SPS) device mainly consisting of a chamber, electrodes, graphite spacers, punches, and graphite dies. By applying pressure and pulsed current heating to the powder samples, a bulk sintered body with high relative density can be obtained.}
%\label{SPS_device}
%\end{figure}
%
%Sample preparation SPS-bulk WC and XRD Preparation Method of Measurement Samples:
%
Setup of the SPS is as follows: Firstly, {WC powders with the grain size of nearly 1.0 $\mu$m} provided by KOJUNDO CHEMICAL LABORATORY CO., LTD. to be sintered are placed in the region surrounded by punches and a graphite die [powder area in Fig.~\ref{equipment}(b)]. Next, the chamber is purged with an Ar atmosphere. Finally, sintering is performed at 1700–1900 $^\circ$C {and 10 min} under a maximum pressure of 50 MPa and pulse current heating and  mechanically polishing and grinding are done for the surface on the sintering bodies to obtain measurement samples (polycrystalline) with a surface roughness of Ra 0.001 $\mu$m or less. %\red{We show a detailed of SPS method for sample preparation in supplemantal information.} 
Setup of the HP method is as follows: Powders of the materials to be sintered are placed in the container. Then, they are pressurized at 30 MPa and sintered at a temperature of 1500-1800 $^\circ$C. Finally, the surface of the resulting sample (polycrystalline) is polished. % \blue{and the resulting Ra is \green{XX} $\mu$m.} 

%\subsection{Ab initio calculation}\label{Ab initio calculation}
%To verify the reliability of experimental spectra with the developed optical measurement system, 
{We perform {\em ab initio} density functional calculations with Quantum Espresso~\cite{giannozzi_2020} with the Perdew-Burke-Ernzerhof type~\cite{perdew1996phys} for the exchange-correlation functional and norm-conserving pseudopotentioals with the code ONCVPSP (Optimized Norm-Conserving Vanderbilt PSeudopotential \cite{hamann2013optimized}) obtained from the PseudoDojo \cite{van2018pseudodojo}.} We use a 32$\times$32$\times$32 Monkhorst–Pack $k$-mesh for the Brillouin zone integration. The kinetic energy cutoff is set to be 96 Ry for the wave functions and 384 Ry for the charge density. The Fermi energy is estimated with the Gaussian smearing techniques with the width of 0.001 Ry~\cite{methfessel1989high}. {The crystal structures} are fully optimized. %The resulting lattice parameteres are listed in Table[] and are in good agreement with the experimental results[Ref].

{\em Ab initio} calculations for reflectance spectra are performed with RESPACK \cite{Nakamura_2010,nakamura2016ab,nakamura2021respack}: The energy cutoff for the dielectric function is set to 10 Ry. The total number of bands used in the polarization calculation are set to cover unoccupied states up to 40 eV above the Fermi level. The integral over the Brillouin zone is calculated with the generalized tetrahedron technique \cite{fujiwara2003generalization} with smearing of 0.01 eV. For materials containing heavy elements (Au and WC), we consider the spin-orbit interactions \cite{Charlebois_2021}. The reflectance spectra are calculated from 
\begin{eqnarray}
 R(\omega) = \Biggl| \frac{ 1-\sqrt{\epsilon(\omega)} }{ 1+\sqrt{\epsilon(\omega)} } \Biggr|^2
\label{eq:reflectance} 
\end{eqnarray}
with $\epsilon(\omega)$ being a dielectric function in the random phase approximation based on the Lindhard formula \cite{DRAXL_2006}.

{Tests for the developed optical measurement system are performed, given in Supplemental information (SI), where we check spectral reflectances of Al, Au, and Cu in a face-centered cubic (fcc) form. We found that the measured results are in good agreements with previous studies~\cite{PFLUGER1997293} and {\em ab initio} calculations.}

\red

%\subsection{WC}\label{WC}
{Now we consider sample and spectral properties of WC in a hexagonal closed pack (hcp) structure}. Table~\ref{DS_and_DL_parameter} summarize sample features, where the purity is high enough and the density is 15.51 g/cm$^3$ being close to the theoretical value is 15.63 g/cm$^3$. Figure~\ref{WC_SPS} summarizes our results for the WC; the panel (a) is a microscopic image and photo of the SPS sample after polishing. The tiny black dots are the pores on the sample surface. The panel (b) is the X-ray diffraction (XRD) patterns of the sample, where the blue and red lines are observed data and database [ICSD (Inorganic Crystal Structure Database): 77566], respectively. The panel (c) is measured reflectance spectrum (red-solid line), compared with an {\em ab initio} spectrum (blue-solid line) with smearing of 0.01 eV. {The measurement range is the solar energy range, so the range of 0.1 to 2.5 eV (the wavelength of 0.5 to 10.0 $\mu$m) is targeted. In the figure, the bright orange shadow represents the sunlight range (0.6 to 4.0 eV), and the gray shadow represents the principal radiation range (0.12 to 0.6 eV).} We see a clear plasma edge around 0.6 eV (2.0 $\mu$m) corresponding to the cutoff energy (wevelength) of the sunlight. This property is important for suppressing thermal radiation of the absorbed solar energy; the WC has an important advantage as a solar selective absorber. More carefully looking around the low-energy (0.1-0.5 eV), we see a lowering of the measured reflectance compared with the theoretical spectrum, which would be due to the multiple scattering in the pores of the surface. 

To investigate this point, we calculated an {\em ab initio} reflection spectra with smearing of 0.1 eV, denoted by the blue-dashed line, where the smearing size qualitatively represents the strength of scattering by impurities. As can be seen by comparing the result with the smearing factor of 0.01 eV, increasing the smearing broadens the spectrum. Regarding the behavior in the low-energy region, the result with 0.1 eV seems to be in better agreement with experiment. {The spectral reflectance up to 0.6 eV measured by FT-IR (JASCO, FT/IR-4100 with reflectance measurement attachment) is also shown in Fig.~\ref{WC_SPS}(c) with black-dashed line. The result is consistent with our device result. Figure of merit for photothermal conversion of WC is {estimated as} 0.51-0.56 ({\em ab initio}) and 0.37-0.40 (SPS). {Detailed} discussions are given in SI.} We note that there are reports on the synthesis of single crystals of WC \cite{Lundstrom_1984,He_2017,Ma_2018,Hou_2019}. This would be important for improving the quantitative accuracy of the spectral property, which is left to be explored. %In general, reflectance measurements become difficult for the small sample. Then, if we prioritize the measurement, we have to prepare a large sample, but it is difficult to synthesize large-size single crystals. The present study uses the SPS method as a compromise. With SPS, it is possible to prepare the large-size sample and provide a relatively clean sample; it minimizes the number of pores in the sintered body and can withstand the quantitative nature of the reflectance measurement.
\begin{figure}[b] %[tb]
\centering 
\includegraphics[width=0.55\textwidth]{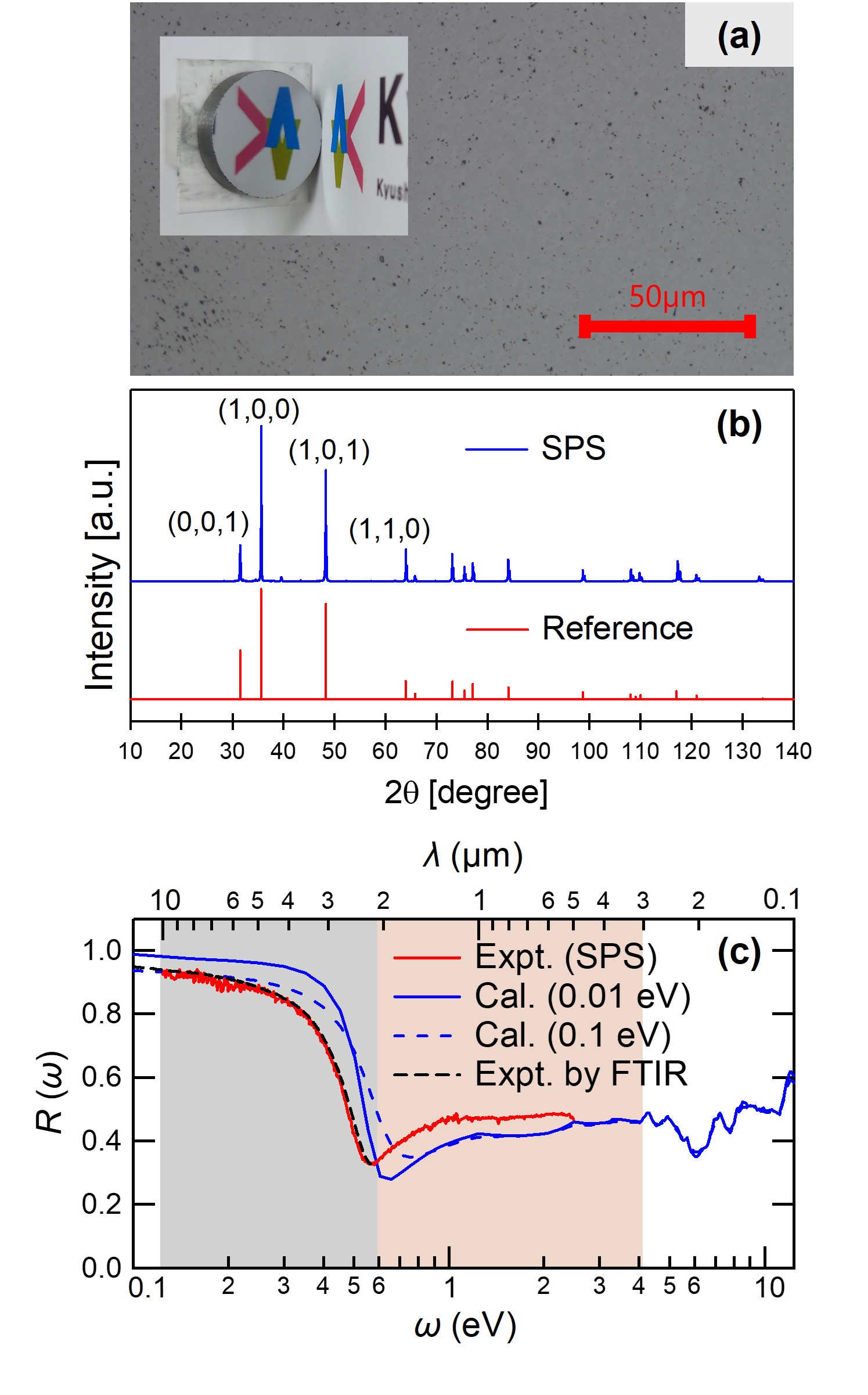}
\vspace{-0.5cm} 
\caption{Results for microcrystalline WC synthesized by the SPS method. (a) Microscopic image and photo of WC. The sample size is 20 mm in diameter and 5 mm in thickness. Small black dots are pores on the surface. (b) XRD patterns of the samples, where blue lines represent the observed data and red lines are database (ICSD: 77566). (c) Experimental reflectance spectrum (red-solid line) compared with an {\em ab initio} result (blue-solid line) considering spin-orbit interactions and smearing of 0.01 eV. We also show an {\em ab initio} result (blue-dashed line) with smearing of 0.1 eV and FT-IR reflectance spectrum (black-dashed line). {In the figure, the bright orange shadow (0.6 to 4.0 eV) represents the spectral range of sunlight, and the gray shadow (0.12 to 0.6 eV) represents the range of radiation.}}
\label{WC_SPS}
\end{figure}
\begin{table}[h!] 
\caption{Purity and density of obtained sintered bodies for WC, TiC, and TiN. The theoretical density is 15.63 g/cm$^3$ for WC, 4.93 g/cm$^3$ for TiC, and 5.43 g/cm$^3$ for TiN. Melting temperature $T_{{\rm m}}$ of each material is also given. The units of purity, density, and $T_{{\rm m}}$ are wt\%, g/cm$^3$, and K, respectively. For TiC and TiN, spark-plasma-sintering (SPS) and hot-pressing (HP) results are compared.} %Lattice parameters based on the X-ray diffraction data are given in \AA, compared with theoretical {\em ab initio} optimized parameters.}
\begin{center} 
\begin{tabular}{l@{\ \ \ }c@{\ \ \ }c@{\ \ \ }c@{\ \ \ }c@{\ \ \ }c@{\ \ \ }c@{\ \ \ }c} \hline \hline \\ [-8pt] %[-20][-5pt] for double column  
     & WC (hcp) & & \multicolumn{2}{c}{TiC (fcc)}  & & \multicolumn{2}{c}{TiN (fcc)}  \\ 
     \cmidrule{2-2} \cmidrule{4-5} \cmidrule{7-8}
     & SPS & & HP & SPS & & HP & SPS \\ \hline \\ [-8pt] %[-20][-5pt] for double column  
     Purity                  & 99.996& & 99.650 & 99.651 & & 99.554 & 99.898 \\ [+5pt]
     Density                 & 15.51 & & 4.43   & 4.86   & & 4.63   & 5.15   \\ [+5pt]
     $T_{{\rm m}}$ & 3143~\cite{handbook_refractory}  & & 
     \multicolumn{2}{c}{3203~\cite{sigalas2000}} & & 
     \multicolumn{2}{c}{3338~\cite{sigalas2000}} \\ [+5pt]
%     $a_{{\rm expt.}}$  & 2.906 & & 4.325  & 4.325       & & 4.243  & 4.242       \\ [+5pt]
%     $c_{{\rm expt.}}$  & 2.837 & & \multicolumn{2}{c}{-}     & & \multicolumn{2}{c}{-}     \\ [+5pt]
%     $a_{{\rm theo.}}$  & 2.922 & & \multicolumn{2}{c}{4.333} & & \multicolumn{2}{c}{4.247} \\ [+5pt]
%     $c_{{\rm theo.}}$  & 2.847 & & \multicolumn{2}{c}{-}     & & \multicolumn{2}{c}{-}     
     \hline \hline 
\end{tabular}
\end{center}
\label{DS_and_DL_parameter} 
\end{table}

%\subsection{TiC and TiN}\label{TiC and TiN}
We now discuss effects of the sintered-body quality on spectral reflectance. For this purpose, we prepare two samples: One is the SPS sample, and the other is the HP one. We consider TiC and TiN in an fcc structure, which are also main components of the TiCN-based cermet and good solar selective absorvers candidates \cite{roux1982optical}. Similarly to WC, these have rather high melting points in Table~\ref{DS_and_DL_parameter}. From here, we write TiC and TiN samples synthesized with the SPS method as TiC (SPS) and TiN (SPS), and samples with the HP method as TiC (HP) and TiN (HP). Table~\ref{DS_and_DL_parameter} summarizes the sample features. The density of TiC (HP) is 89.9\% of the theoretical density, and that of TiC (SPS) is 98.6\%. Similarly, the density of TiN (HP) is 85.3\% of the theoretical density, and that of TiN (SPS) is 94.8\%. Thus, it is clear that the SPS method can generate sintered bodies with few pores. {XRD data {and lattice parameters of} the TiC and TiN samples are shown in SI}.

The upper part of Fig.~\ref{SPS_vs_HP} compares microscopic images of the SPS and HP samples, %\red{
where the panels (a) and (b) are the images of TiC (SPS) and TiC (HP), respectively, and (c) and (d) are the results for TiN (SPS) and TiN (HP), respectively. The SPS samples clearly have smaller pores, being consistent with the density data of Table~\ref{DS_and_DL_parameter}. However, the SPS sample still includes many small pores and therefore the reflection intensity would be quantitatively affected. %} 

\clearpage
The lower part of Fig.~\ref{SPS_vs_HP} compares measured reflectance spectra of (e) TiC (SPS), (f) TiC (HP), (g) TiN (SPS), and (h) TiN (HP). Red lines are our spectra, green solid and dashed lines are other experimental spectra of films~\cite{PFLUGER1997293, roux1982optical}, and blue lines are {\em ab initio} spectra. The agreement between the SPS and theoretical results is reasonable, but, in the low energy region (0.1-2 eV), the experimental reflectance is smaller than the {\em ab initio} one. %This is a similar situation to the WC and would be due to absorption due to the multiple scattering occurring in the pores in the sintered body. 
From a comparison between the SPS and HP samples, we see a clear difference. As can be seen from the microscopic images in Fig.~\ref{SPS_vs_HP}, the pore size is rather large in the HP sample compared to the SPS sample, resulting in a significant decrease in reflectance due to the multiple scattering in the HP sample. {More detailed analyses for the porosity and its effect on the spectral reflectance are given in SI.}
\begin{figure}[!] 
   \begin{center} 
   \includegraphics[width=1\textwidth]{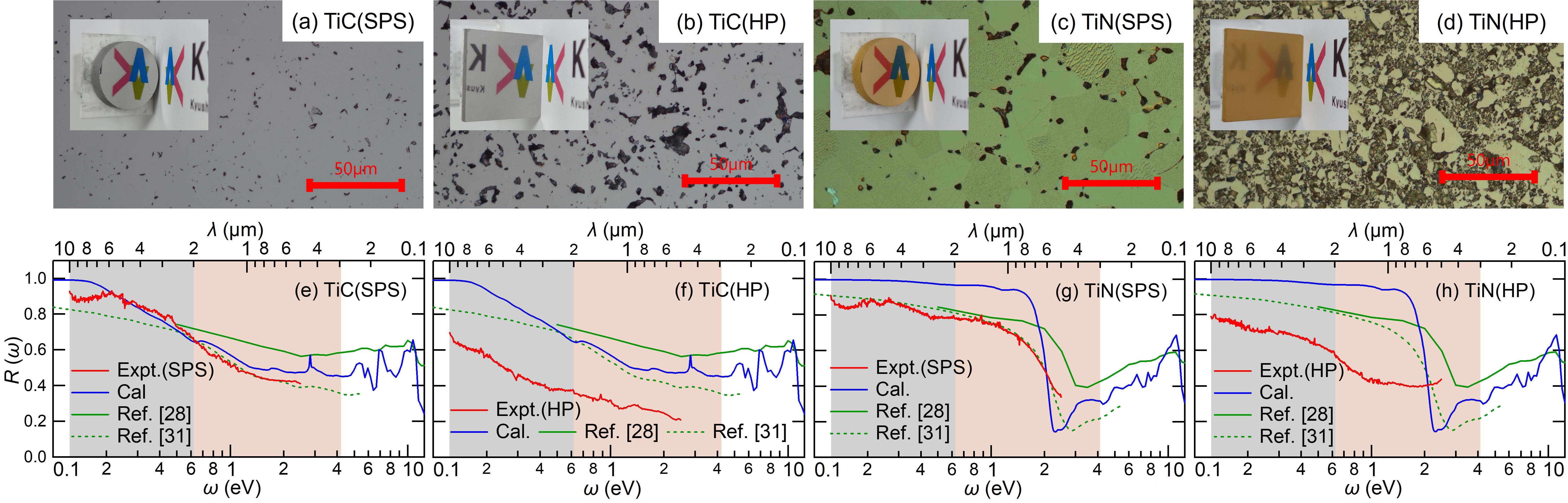}%width=0.34
   \end{center} 
   %\vspace{-0.5cm} 
   \caption{Microscopic images of (a) TiC (SPS), (b) TiC (HP), (c) TiN (SPS) and (d) TiN (HP). The sample size of the SPS is the diameter of 20 mm and the thickness of 5 mm, and the HP sample size is 25$\times$25$\times$2 mm. Black spots are pores on the surface. Insets show sample photos after polishing. Reflectance spectra of (e) TiC (SPS), (f) TiC (HP), (g) TiN (SPS), and (h) TiN (HP). Red lines are measured spectra with our developed optical measurement system, green solid and dashed lines are other experimental spectra of films~\cite{PFLUGER1997293, roux1982optical}, and blue lines are {\em ab initio} spectra with smearing of 0.01 eV.}
   \label{SPS_vs_HP} 
\end{figure}
%\begin{figure}[h!] 
%\begin{center} 
%\includegraphics[width=0.5\textwidth]{Fig-7_TiCTiN_Reflectivity.jpg}
%\includegraphics[width=0.4\textwidth]{Fig-7_TiCTiN_Reflectivity.jpg}
%\end{center} 
%\vspace{-0.5cm} 
%\caption{%measured reflectivity spectra of (a)WC(SPS), (b)TiC(SPS), (c)TiC(Hot press), (d)TiN(SPS) and (e)TiN(Hot press) as a function of photon energy omega or photon wavelength lamda (upper scale). The measurement curves are given by red solid curves and compared with the theoritical curves (blue solid curves) and privious experimental results (green and --- solid curves) for TiC(cite) and TiN(cite).}
%Reflectance spectra of (a) TiC (SPS), (b) TiC (HP), (c) TiN (SPS), and (d) TiN (HP). Red lines are the spectra with our developed optical measurement system, and green solid and dashed lines are other experimental spectra of films.\cite{PFLUGER1997293, roux1982optical}. Also, blue lines are {\em ab initio} spectra.}
%\label{TiC-TiN-reflecatance}
%\end{figure}

%\clearpage

%\section{Conclusion}\label{Conclusion}
In the present study, we have reported on the first experimental reflectance spectrum of tungsten carbide WC. We have developed a measurement system for the reflectance spectrum in visible to mid-infrared region. The microcrystalline WC sample has been prepared by sintering the powders with the SPS technique. The resulting reflection spectrum of WC exhibits a sharp low-energy plasma edge around $\sim 0.6$ eV (2.0 $\mu$m) corresponding to the cutoff energy of the sunlight, thus suppressing the thermal radiation and proving it a highly promising ingredient of the solar selective absorber. The measured spectrum is in good quantitative agreement with {\em ab initio} calculations. 

In addition, we have examined the effect of the sintered-sample quality on the reflectance spectrum by comparing the samples synthesized by the SPS and HP method. Effectiveness of the SPS method %\red{
prior
%} 
to the HP method has been proved; the HP sample exhibits a rather small reflectance intensity, while the SPS can well reduce the pore size and density in the sample, resulting in a reasonable agreement with the previous experimental and {\em ab initio} spectra.
%because of the absorption due the the multiple scattering coming from the high pore density of the sample. On the other hand, 

%To further improve the quantitative accuracy, it would be necessary to perform experiments using single crystals beyond the sintered body.

%\section{Acknowledgments}
%\label{acknowledgments}
\vspace{0.2cm} 
\noindent \textbf{Acknowlegements}

\vspace{0.2cm}
\noindent The authors acknowledge MARUWAGIKEN Co., Ltd. for mechanically polishing of the HP samples. This research was supported by JSPS KAKENHI Grant Numbers JP19K03673, JP22H01183, JP23H01353, and JP23H01126.
%\bibliography{Paper}% Produces the bibliography via BibTeX.

%\begin{thebibliography}{99}
\newcommand{\newblock}{}

\end{document}

\title[Supplemental Information]{ Supplemental Information for Reflectance spectral studies of spark plasma sintered tungsten carbide pellet}

\author{Toshiharu Chono$^{1}$, Hisashi Tokutomi$^{2}$, Kazuma Nakamura$^{1,3}$\thanks{E-mail: kazuma@mns.kyutech.ac.jp} \& Koji Miyazaki$^{1,3,4}$\thanks{E-mail: miyazaki.koji.962@m.kyushu-u.ac.jp}}

\address{$^{1}$Graduate School of Engineering, Kyushu Institute of Technology, 1-1 Sensui-cho, Tobata-ku, Kitakyushu, Fukuoka 804-8550, Japan \\
$^{2}$KOJUNDO CHEMICAL LABORATORY CO., LTD., 5-1-28, Chiyoda, Sakado, Saitama 350-0284, Japan \\
$^{3}$Integrated Research Center for Energy and Environment Advanced Technology, Kyushu Institute of Technology, 1-1 Sensui-cho, Tobata-ku, Kitakyushu, Fukuoka 804-8550, Japan \\
$^{4}$Graduate School of Engineering, Kyushu University, 744 Motooka, Nishi-ku, Fukuoka 819-0395, Fukuoka, Japan}
\ead{kazuma@mns.kyutech.ac.jp, miyazaki.koji.962@m.kyushu-u.ac.jp}
\vspace{10pt}
\begin{indented}
%\item[]August 2024
 \item[] June 2024
\end{indented}

\section{Measurement details and benchmark test}\label{Benchmark}
Here, we describe measurement details to confirm the quantitative validity of our developed optical measurement system. As benchmark samples, the metal thin films were prepared by a vacuum thermal evaporation method (ULVAC VPC-260) {\cite{TAKASHIRI2007246}}. With the residual gas pressure with the order of 10$^{-5}$ Torr, an electric current of 60-70 A is applied to a tungsten heater for metal evaporation, and optically thick films with more than 200 nm thickness were deposited on glass substrates.

The reflectance spectrum of the measured sample $R_{sample}$ is corrected using a reflectance spectrum of a calibration sample $R_{ref}$ as  
\begin{eqnarray}
 R_{sample}(\lambda) = \frac{I_{sample}(\lambda)}{I_{ref}(\lambda)} R_{ref}(\lambda), 
\label{eq:reflectance} 
\end{eqnarray}
where $I_{sample}$ and $I_{ref}$ are reflection intensities of the measured and calibration samples, respectively, and the calibration sample is an Ag thin film. 

\begin{figure}[h!] 
\begin{center} 
\includegraphics[width=0.55\textwidth]{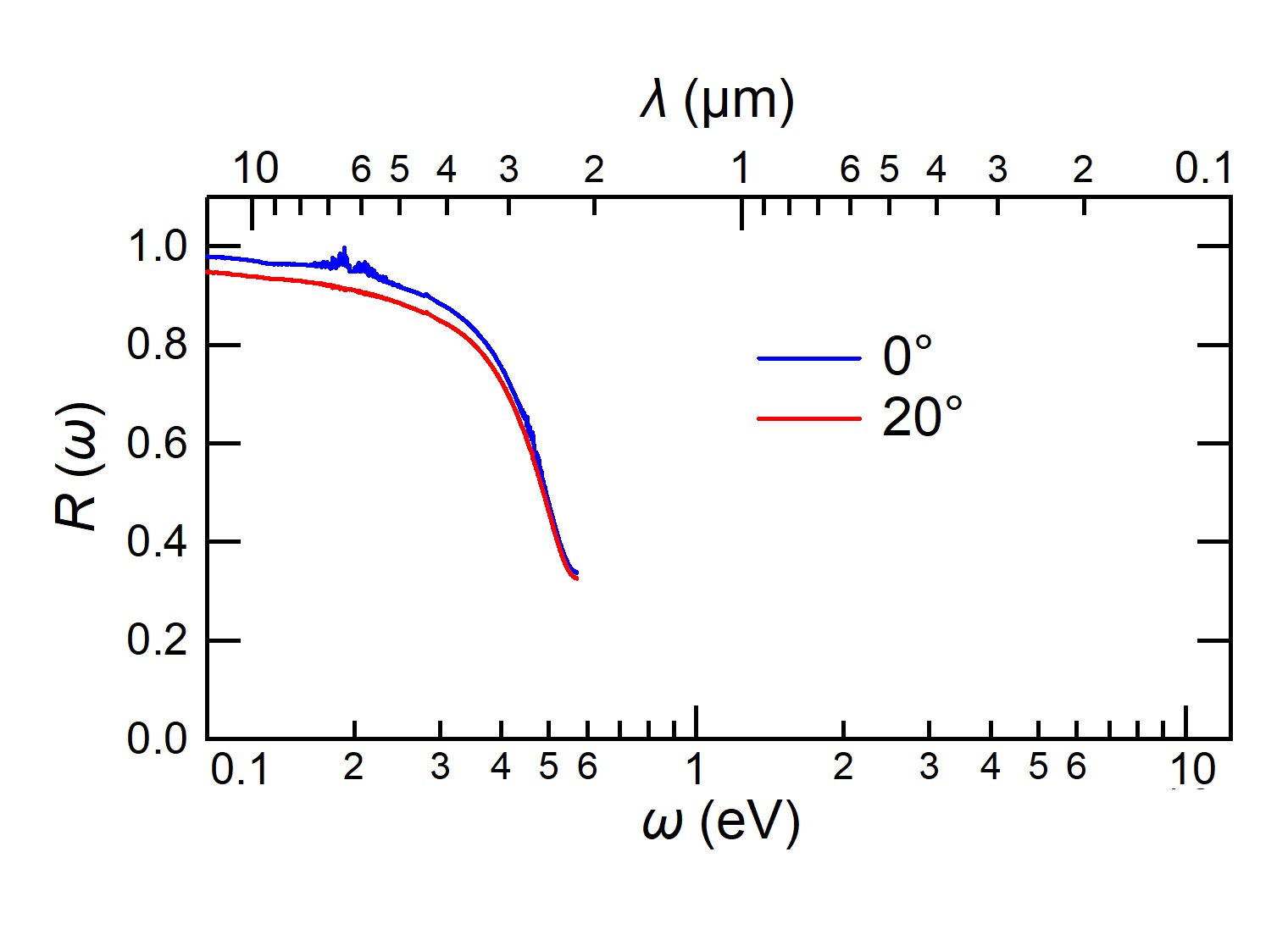}
\vspace{-0.5cm} 
\end{center} 
\caption{Incident angle dependence of FT-IR spectral reflectance, where blue and red curves are the results of the 0$^{\circ}$ and 20$^{\circ}$ settings, respectively.}
\label{reflectivity_angle}
\end{figure}

In the actual measurement [Fig. 1(a) of the main text], the sample is tilted 20$^{\circ}$ and the light reflected specularly at an incident angle of 20$^{\circ}$ is measured. Since it is a sufficiently thick mirror sample, we are able to measure a reflection spectrum that is hardly different from that measured with perpendicular incidence. To check the angle dependence of the reflection spectrum, we show in Fig. S~\ref{reflectivity_angle} the spectral reflectance of WC with the FT-IR measurements, where we see that there is no appreciable difference between the 0$^{\circ}$ (blue curve) and 20$^{\circ}$ (red curve) settings.

\begin{figure}[h!] 
\begin{center} 
\includegraphics[width=0.55\textwidth]{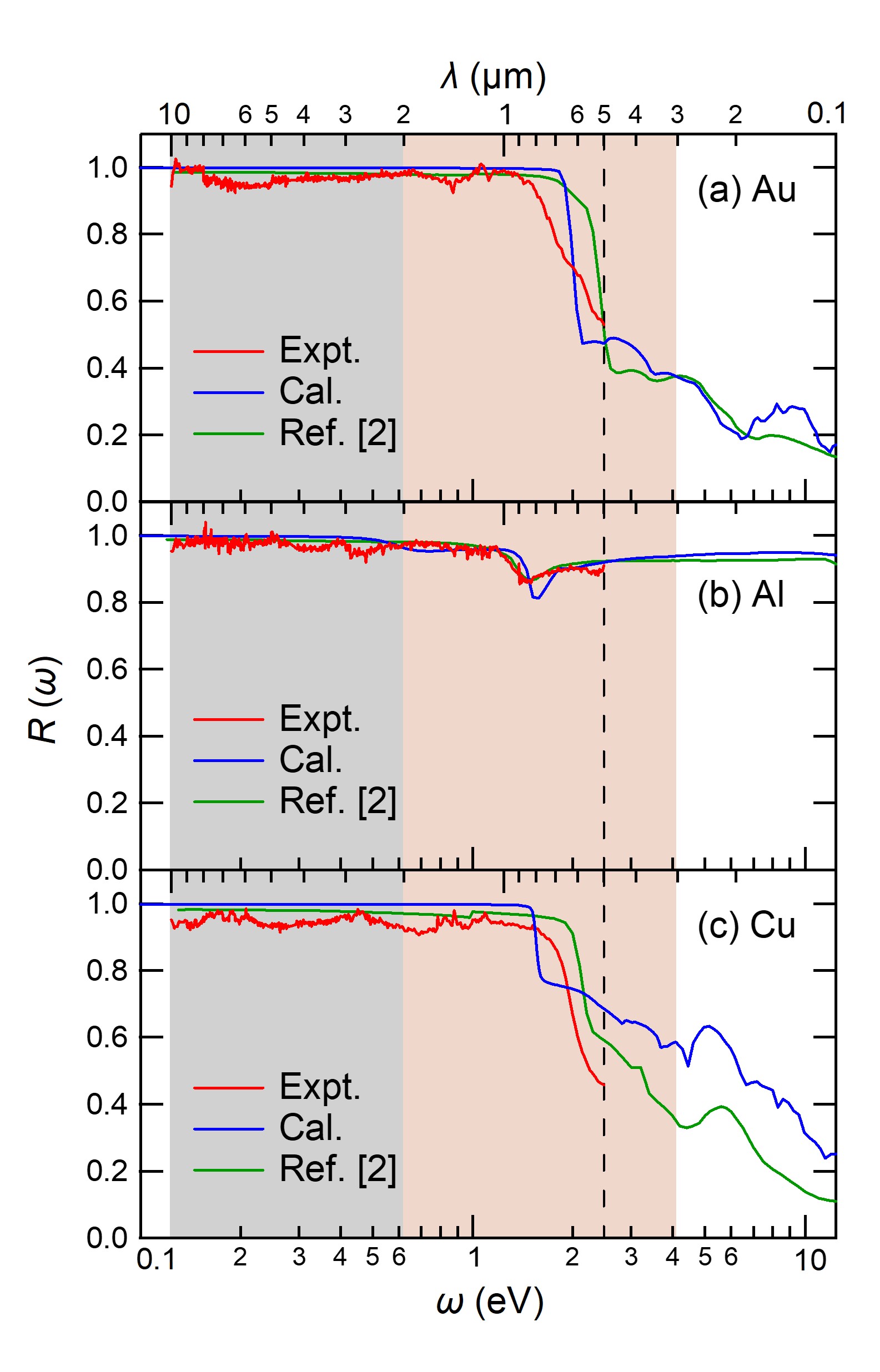}
\end{center} 
\vspace{-0.5cm} 
\caption{{Reflectance spectra of (a) Au, (b) Al, and (c) Cu. The red lines are measured spectra for the metallic films and green lines are other experimental film spectra taken from Ref.~\cite{PFLUGER1997293}. Also, blue lines are our calculated {\em ab initio} spectra with the fcc structure. In the figure, the bright orange shadow (0.6 to 4.0 eV) represents the spectral range of sunlight, and the gray shadow (0.12 to 0.6 eV) represents the range of radiation. The vertical dashed line represents the upper bound of our developed optical measurement system. Note that we do not display the CO$_2$ absorption band around 0.3 eV.}}
\label{Metal}
\end{figure}

{Figures S~\ref{Metal} (a), (b), and (c) shows measured reflectance spectra for benchmark metals of Au, Al, and Cu, respectively. Our results (red lines) are compared with other experimental film results (green lines) and {\em ab initio} calculations (blue lines) with a face-centered cubic (fcc) structure. We see a good agreement among them. The measurement range of our interest is the solar energy range, so the range of 0.1 to 2.5 eV (the wavelength of 0.5 to 10.0 $\mu$m) is targeted. In the figures, the bright orange shadow represents the spectral range of sunlight (0.6 to 4.0 eV), and the gray shadow represents the principal radiation range (0.12 to 0.6 eV). These shaded regions are important in discussing the solar selective  absorption performance.}% being desirable to absorb sunlight well and suppress radiation.

\section{Figure of merit for photothermal conversion}

Figure of merit $\eta_{{\rm FOM}}(T)$ for photothermal conversion at a temperature $T$ is given as a ratio of a usable heat $Q_H(T)$ to the input solar energy $cI_0$ as
\begin{eqnarray}
    \eta_{{\rm FOM}}(T) \equiv \frac{Q_H(T)}{cI_0}, 
    \label{equation_eta}
\end{eqnarray}
where $c$ and $I_0$ are the solar concentration and the solar flux intensity, respectively. The $Q_H(T)$ is written as~\cite{bermel2012selective}  
\begin{eqnarray}
    Q_H(T) = B \alpha_s c I_0 - \epsilon_t(T) \sigma T^4,   
    \label{QH} 
\end{eqnarray}
where the first term of the right-hand side describes the input heat obtained by light absorption, and the second term represents the heat loss due to the thermal radiation. $B$ is the transmittance of a glass tube enclosing the solar collector, and $\sigma$ is the Stefan–Boltzmann constant. $\alpha_s$ is the solar absorptivity given as the wavelength integral as~\cite{Cao_2014,bermel2012selective,zhang2022solar}
\begin{equation}
\alpha_s = \frac{\int_{\lambda_l}^{\lambda_h} \bigl(1-R(\lambda)\bigr) I_{sol}(\lambda) d\lambda}{ \int_{\lambda_l}^{\lambda_h} I_{sol}(\lambda) d\lambda}, 
\label{equation_alp}
\end{equation}
where $R(\lambda)$ is the reflectance spectra as a function the wavelength $\lambda$. $I_{sol}(\lambda)$ is the solar spectral irradiance (air mass of 1.5) taken from \cite{GUEYMARD2001325}. $\lambda_l$ and $\lambda_h$ are the lower and higher cutoff wavelengths, respectively, and these are set to 0.28 $\mu$m and 4 $\mu$m in the present study. Similarly, the thermal emissivity $\epsilon_t(T)$ at the temperature $T$ is written as~\cite{Cao_2014,bermel2012selective,zhang2022solar}
\begin{equation}
\epsilon_t(T) = \frac{\int_{\lambda_L}^{\lambda_H} \bigl( 1-R(\lambda) \bigr) I_{b}(T, \lambda) d\lambda}{\int_{\lambda_L}^{\lambda_H} I_{b}(T, \lambda) d\lambda}, 
\label{equation_eps}
\end{equation}
where $I_b (T,\lambda)$ is the spectral blackbody radiative intensity having the temperature dependence, which is taken from \cite{Max-plank}. $\lambda_L$ and $\lambda_H$ are the lower and higher cutoff wavelengths for the emissivity evaluation, respectively, and are set to 0.1 $\mu$m and 10 $\mu$m. By inserting Eq.~(\ref{QH}) into Eq.~(\ref{equation_eta}), we obtain the following expression 
\begin{eqnarray}
    \eta_{{\rm FOM}}(T) = B\alpha_s - \frac{\epsilon_t(T) \sigma T^4}{cI_0}.  
    \label{eta}
\end{eqnarray}

%Experiment : 0.1 $\mu$m and 10 $\mu$m using fitting, 
%Calculation : 0.1 $\mu$m and 124 $\mu$m
In the present calculation, we calculate $\eta_{{\rm FOM}}(T)$ in Eq.~(\ref{eta}) for two conditions. The first is a condition with the linear fresnel reflector solar thermal power generation~\cite{Concentrating Solar Power}, where we set $T$ to 673 K, $c$ to 30 suns, $I_0$ to 863 W/m$^2$, and $B$ to 0.91~\cite{Zhang_1999}. The other is based on solar thermoelectric generator and solar water heater ~\cite{solar_water_heater,solar_thermoelectric}, where we set $T$ to 400 K, $c$ to 1 suns (unconcentrated), $I_0$ to 1 kW/m$^2$, and $B$ to 1 (no glass envelope).

Table S~\ref{tb:arrst} lists our calculated $\eta_{{\rm FOM}}$. From the table, we see that the WC results by the first-principles (FP) calculation give the best performance. Unfortunately, the WC result by the spark plasma sintering (SPS) method does not achieve the performance of the FP calculation. This may probably be due to the fact that the FP calculation shows the ideal performance of a crystal with no pores, while WC(SPS) is a polycrystal (sintered body). This point needs further improvements in the future.
\begin{table}[t]
 \caption{Summary of figure of merit $\eta_{{\rm FOM}}$ for photothermal conversion in Eq.~(\ref{eta}). The calculated solar absorptivity $\alpha_s$ in Eq.~(\ref{equation_alp}) and thermal emissivity $\epsilon_t$ in Eq.~(\ref{equation_eps}) are also shown.  The condition 1 is $T=673$ K, $c=30$ suns, $I_0=863$ W/m$^2$, and $B=0.91$~\cite{Zhang_1999}. The condition 2 is $T=400$ K, $c=1$ suns, $I_0=1$ kW/m$^2$, and $B=1$ ~\cite{solar_water_heater,solar_thermoelectric}. The bold in the $\eta_{{\rm FOM}}$ column represent the best data among the lists.}
  \vspace{.25cm} 
  \centering
  \begin{tabular}{|c@{\ \ \ \ }|c@{\ \ \ \ }|c@{\ \ \ \ }|c@{\ \ \ \ }|c@{\ \ \ \ }|c@{\ \ \ \ }|c|} \hline  
    Condition & \multicolumn{3}{c|}{condition 1} & \multicolumn{3}{c|}{condition 2} \\ \hline 
    Material  & $\alpha_s$ & $\epsilon_t$ & $\eta_{{\rm FOM}}$  & $\alpha_s$ & $\epsilon_t$ & $\eta_{{\rm FOM}}$ \\ \hline 
    WC (SPS)  & 0.515 & 0.160 & 0.397 & 0.515 & 0.103 & 0.366 \\ \hline \hline 
    WC$_{E\parallel x}$ (FP) & 0.577 & 0.038 & 0.508 & 0.577 & 0.012 & 0.560  \\ \hline 
    WC$_{E\parallel z}$ (FP) & 0.524 & 0.023 & 0.467 & 0.524 & 0.011 & 0.509  \\ \hline \hline 
    TiC (HP)  & 0.721 & 0.481 & 0.439 & 0.721 & 0.423 & 0.106  \\ \hline 
    TiC (SPS)  & 0.538 & 0.138 & 0.428 & 0.538 & 0.113 & 0.374  \\ \hline 
    TiC (FP) & 0.487 & 0.117 & 0.390 & 0.487 & 0.058 & 0.403  \\ \hline 
    TiC~\cite{roux1982optical} & 0.546 & 0.229 & 0.393 & 0.546 & 0.207 & 0.245  \\ \hline 
    TiC~\cite{palik} & 0.385 & 0.188 & 0.265 & 0.385 & 0.163 & 0.148  \\ \hline \hline 
    TiN (HP)  & 0.571 & 0.285 & 0.391 & 0.571 & 0.256 & 0.198  \\ \hline 
    TiN (SPS)  & 0.445 & 0.150 & 0.338 & 0.445 & 0.130 & 0.256  \\ \hline 
    TiN (FP) & 0.339 & 0.016 & 0.301 & 0.339 & 0.014 & 0.318  \\ \hline 
    TiN~\cite{roux1982optical} & 0.457 & 0.134 & 0.356 & 0.457 & 0.120 & 0.283  \\ \hline 
    TiN~\cite{palik} & 0.302 & 0.100 & 0.229 & 0.302 & 0.083 & 0.181  \\ \hline 
  \end{tabular}
  \label{tb:arrst}
\end{table}

\begin{figure}[htb] 
\begin{center} 
\includegraphics[width=0.5\textwidth]{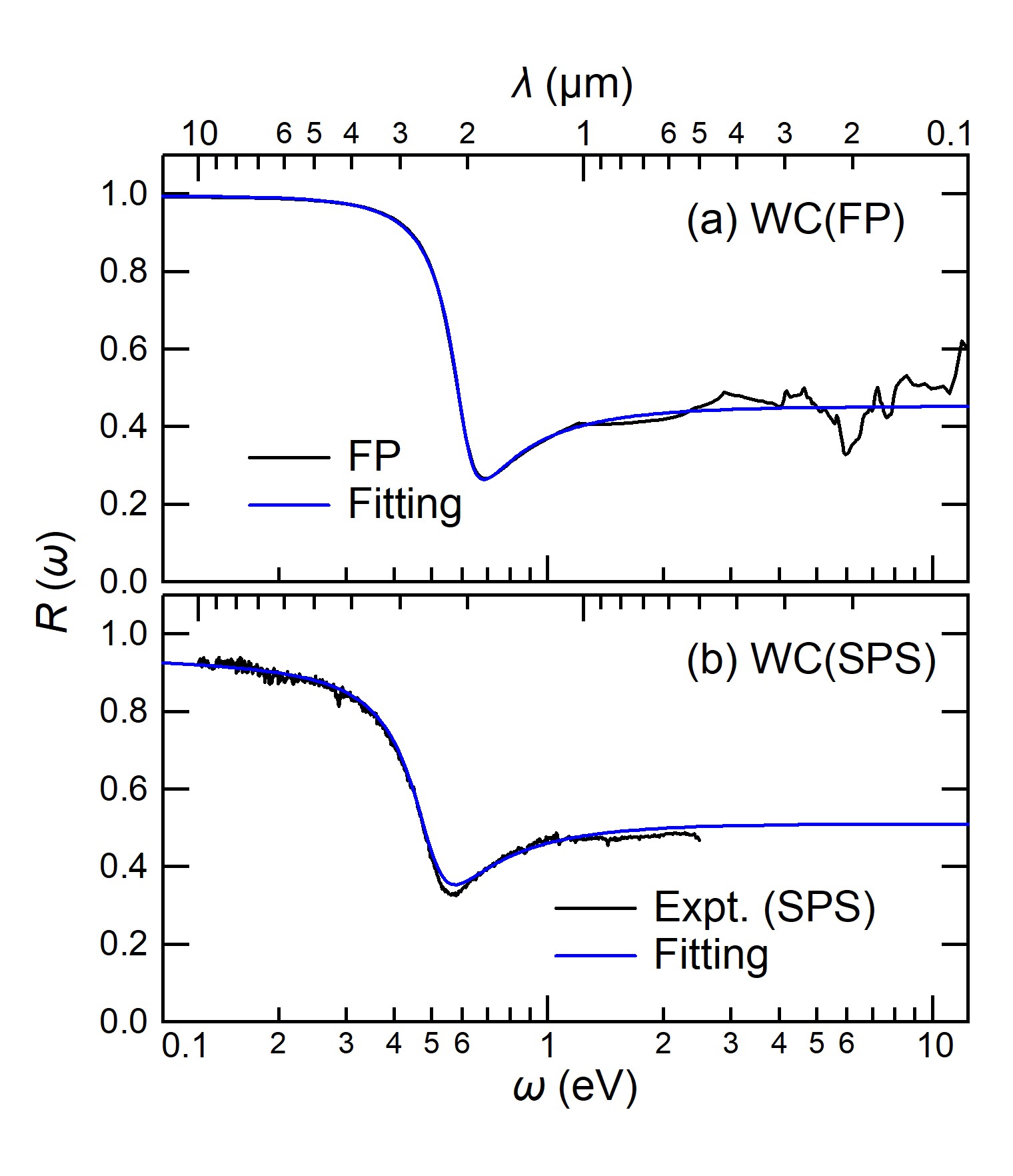}
\vspace{-0.5cm} 
\end{center} 
\caption{Fitting of the DL model in Eq.~(\ref{DL-model}) to (a) first principles (FP) and (b) experimental data of WC. The blue and black curves represent the spectral reflectance of the DL model and data, respectively.}
\label{fitting_WC}
\end{figure}

\begin{figure}[htb] 
\begin{center} 
\includegraphics[width=0.7\textwidth]{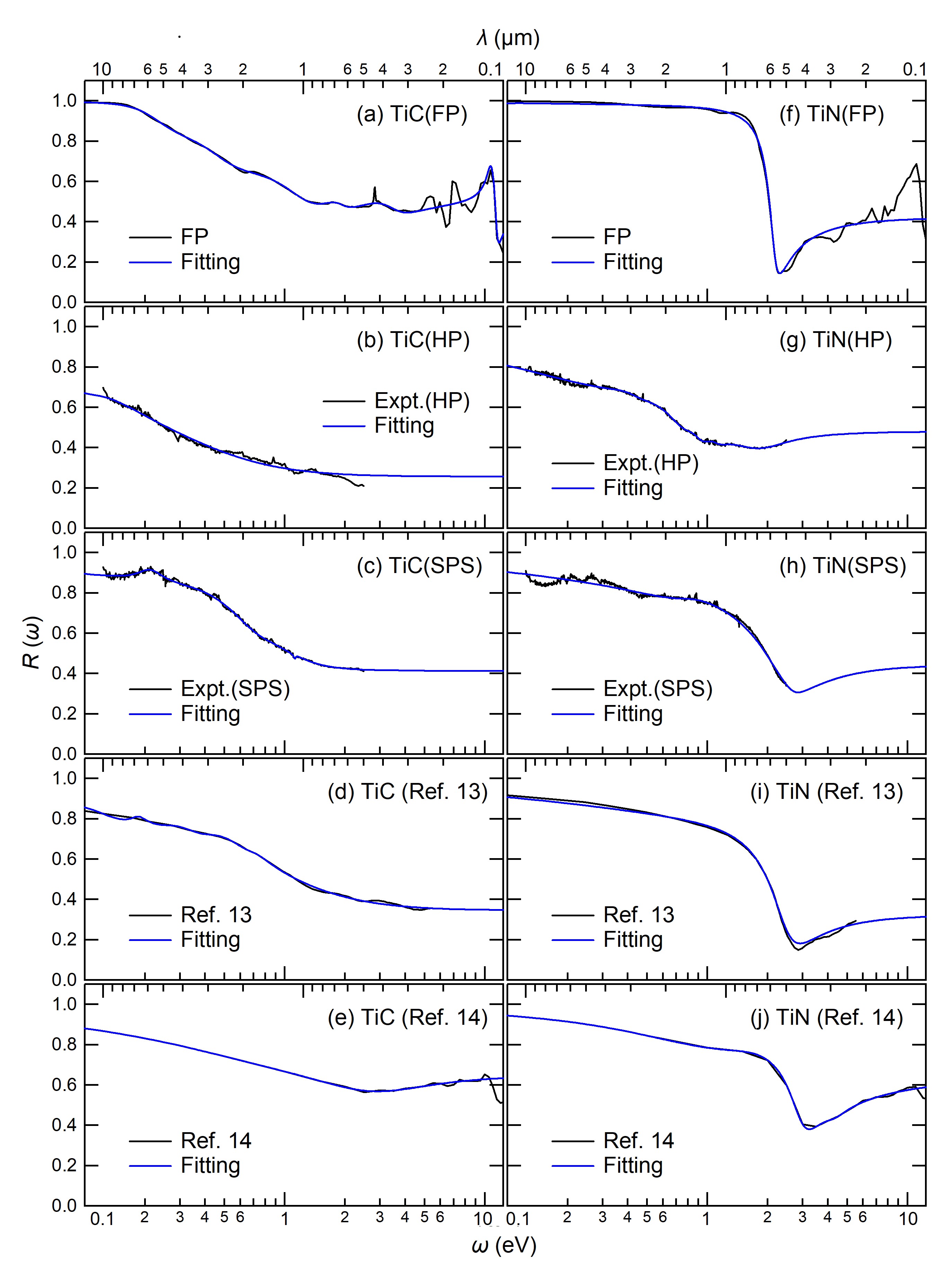}
\vspace{-0.5cm} 
\end{center} 
\caption{Fitting of the DL model in Eq.~(\ref{DL-model}) to FP and experimental data of TiN and TiC. Panels (a)-(e) are TiC, and (f)-(j) are TiN. The blue and black curves represent the spectral reflectance of the DL model and data, respectively. The bottom two panels are the fitting results for other experimental data~\cite{roux1982optical,palik}.}
\label{fitting_TiC_TiN}
\end{figure}

To evaluate of the integral of $\alpha_s$ in Eq.~(\ref{equation_alp}) and $\epsilon_t(T)$ in Eq.~(\ref{equation_eps}), we use the Drude-Lorenz (DL) model~\cite{reffit} as  
\begin{eqnarray}
    \epsilon(\omega)=\epsilon_{\infty} - \frac{\Omega_p^2}{\omega^2+i\omega\Gamma} + \sum_{i}^{M}\frac{\Omega_{pi}^2}{\Omega_i^2-\omega^2-i\omega\Gamma_i}, 
    \label{DL-model}
\end{eqnarray}
where $\Omega_p$ is a model plasma frequency, $\Gamma$ is a linewidth, $\epsilon_{\infty}$ is a parameter due to interband response, $\Omega_i$ is the $i$-th oscillator frequency, $\Omega_{pi}$ is the $i$-th model plasma frequency, and $\Gamma_i$ is the $i$-th linewidth. Also, $M$ is the total number of oscillators, and the $M$ = 5 in the present study. These parameters are determined to reproduce {\em ab initio} and experimental spectral reflectance using the software of \cite{reffit}. Figure S~\ref{fitting_WC} shows the fitting results of WC; (a) the fitting to the {\em ab initio} calculation and (b) the fitting to the experimental result. Blue and black curves describe the DL model in Eq.~(\ref{DL-model}) and the data, respectively. We see a reasonable agreement between the two. The same fittings are also performed for TiC and TiN, and the results are shown in Fig. S~\ref{fitting_TiC_TiN}.

%\clearpage 

\section{Relative density dependence of spectral reflectance}
\begin{figure}[h!] 
\begin{center} 
\includegraphics[width=1\textwidth]{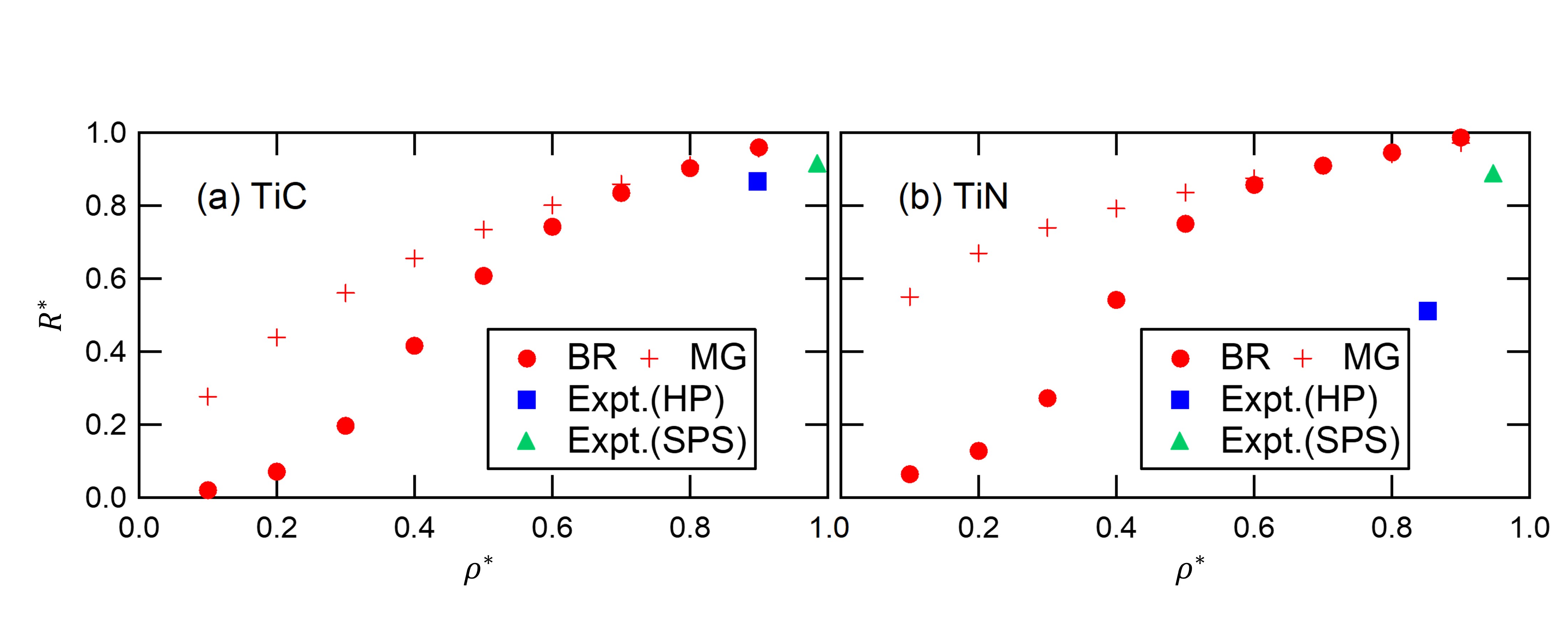}
\vspace{-0.5cm} 
\end{center} 
\caption{Relative integral intensity of the spectral reflectance $R^*$ in Eq.~(\ref{sum_reflectivity_ratio}) as a function of the relative density $\rho^*$ in Eq.~(\ref{density_ratio}) of (a) TiC and (b) TiN. The blue squares indicate the experimental results for the hot-press (HP) samples, and the green triangles indicate the experimental results for the spark-plasma-sintering (SPS) samples. The red dots indicate the simulation results based on the Bruggeman (BR) model in Eq.~(\ref{bruggeman}) and the red crosses indicate the simulation results based on the Maxwell-Garnet (MG) model in Eq.~(\ref{maxwell-garnett}).}
\label{density_vs_reflectivity}
\end{figure}

In this section, we consider a density dependence on the spectral reflectance. To this end, we consider the relative density $\rho^*$ as 
\begin{eqnarray}
    \rho^* = \frac{\rho}{\rho_{ideal}}, 
    \label{density_ratio} 
\end{eqnarray}
where $\rho$ is a density of a sample and $\rho_{ideal}$ is an ideal (theoretical) density. Similarly, a relative integral intensity of the spectral reflectance $R^*$ is introduced as 
\begin{eqnarray}
    R^* = \frac{\int_{\omega_l}^{\omega_h} R(\omega) d\omega}{ \int_{\omega_l}^{\omega_h} R_{ideal}(\omega) d\omega}, 
    \label{sum_reflectivity_ratio} 
\end{eqnarray}
where $R(\omega)$ is a reflectance spectrum of a sample and $R_{ideal}(\omega)$ is an ideal reflectance spectrum, for which we refer to it as {\em ab initio} calculations. The integral domain is set to $\omega_l$ = 0.01 eV and $\omega_h$ = 4.0 eV. Figure S~\ref{density_vs_reflectivity} show the $\rho^*$-$R^*$ plot of (a) TiC and (b) TiN, where squares and triangles represent the hot-pressing (HP) and SPS results, respectively. The relative density dependence of $R^*$ appears to be large for TiN. 

To understand overall profile of the $\rho^*$-$R^*$ plot, we perform model analyses with using the Bruggeman and Maxwell-Garnett models. These models are based on an effective medium theory, and widely used to simulate dielectric property of composite materials like cermet. The formula for the Bruggeman approximation is given as~\cite{bruggeman}
\begin{eqnarray}
    f  \frac{\epsilon_{ideal}(\omega)-\epsilon^{BR}(\omega)}{\epsilon_{ideal}(\omega)+2\epsilon^{BR}(\omega)} 
+ (1-f)\frac{\epsilon_{air}(\omega)-\epsilon^{BR}(\omega)}{\epsilon_{air}(\omega)+2\epsilon^{BR}(\omega)} = 0.
    \label{bruggeman} 
\end{eqnarray}
where $f$ is a volume fraction of the target material TiC or TiN, and $\epsilon_{ideal}(\omega)$ corresponds to a dielectric function of the bulk TiC or TiN, which are taken from the {\em ab initio} data. $\epsilon_{air}(\omega)$ is a dielectric function of air, and $\epsilon^{BR}(\omega)$ is the Bruggeman dielectric function, which is obtained from solving Eq.~(\ref{bruggeman}). The dielectric function $\epsilon^{MG}(\omega)$ based on the Maxwell-Garnett model is the following formula as~\cite{maxwell-garnett}

\begin{eqnarray}
    \epsilon^{MG}(\omega) = \epsilon_{air}(\omega) \frac{\epsilon_{ideal}(\omega)+2\epsilon_{air}(\omega)+2 f (\epsilon_{ideal}(\omega)-\epsilon_{air}(\omega))}{\epsilon_{ideal}(\omega)+2\epsilon_{air}(\omega)- f (\epsilon_{ideal}(\omega)-\epsilon_{air}(\omega))}. 
    \label{maxwell-garnett} 
\end{eqnarray}
By using these models, we calculate the model spectral reflectance $R(\omega)$ and the relative density dependence of $R^*$ in Eq.~(\ref{sum_reflectivity_ratio}).

Figure S~\ref{density_vs_reflectivity} displays our simulation results, where red dots and crosses are the Bruggeman and Maxwell-Garnett models, respectively. The two results appear to be different in the smaller $\rho^*$. For comparing with experimental results, in the case of TiC, the agreement between the simulations and experiments is reasonable. On the other hand, in TiN, the experimental result is rather lower than the simulation results. This difference may be due to the assumption of uniform pore density in the model. 

\section{Lattice parameters of WC, TiC, and TiN}
\begin{figure}[h!] 
\begin{center} 
\includegraphics[width=0.48\textwidth]{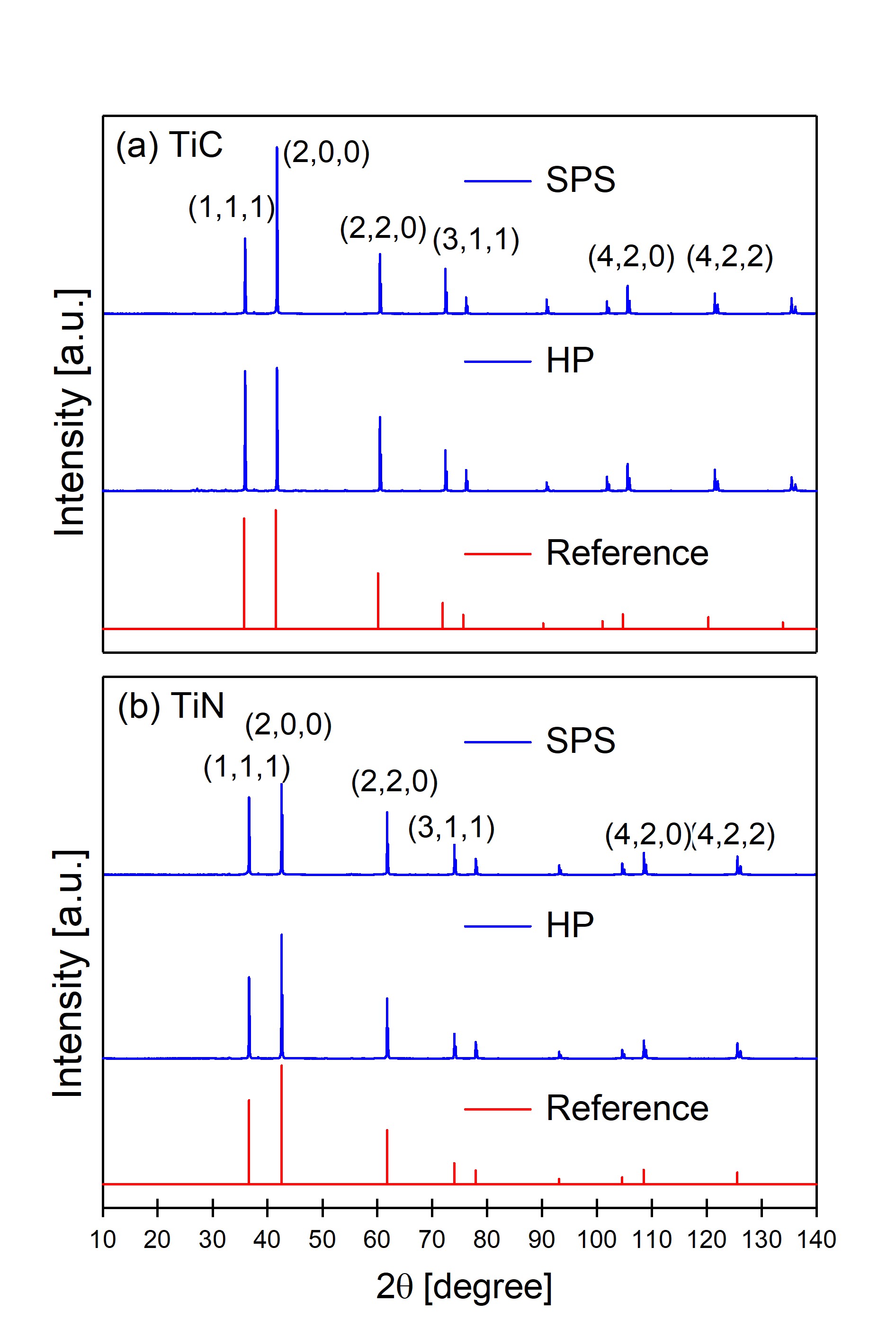}
\vspace{-0.5cm} 
\end{center} 
\caption{XRD patterns of samples obtained with spark-plasma-sintering (SPS) and hot-press (HP) methods. Blue lines are observed data and red lines are database (NIST: 62979 for TiC and ICSD: 64904 for TiN): (a) TiC and (b) TiN.}
\label{XRD_TiCTiN}
\end{figure}

Figure S~\ref{XRD_TiCTiN} shows X-ray diffraction (XRD) patterns of (a) TiC and (b) TiN samples, where the results obtained from the SPS and HP methods are compared. Blue lines are observed data and red lines are database [NIST (The National Institute of Standards and Technology database): 62979 for TiC and ICSD (Inorganic Crystal Structure Database): 64904 for TiN]. From the figure, we do not see a significant difference between the SPS and HP XRD patterns. The resulting lattice parameters are listed in Table S~\ref{lattice_parameter} and are in good agreement with the {\em ab initio} results.
\begin{table}[h!] 
\caption{Lattice parameters based on the X-ray diffraction (XRD) data are given in \AA, compared with {\em ab initio} optimized parameters.}
\begin{center} 
\begin{tabular}{l@{\ \ \ }c@{\ \ \ }c@{\ \ \ }c@{\ \ \ }c@{\ \ \ }c@{\ \ \ }c@{\ \ \ }c} \hline \hline \\ [-8pt] %[-20][-5pt] for double column  
     & WC (hcp) & & \multicolumn{2}{c}{TiC (fcc)}  & & \multicolumn{2}{c}{TiN (fcc)}  \\ 
     \cmidrule{2-2} \cmidrule{4-5} \cmidrule{7-8}
     & SPS & & HP & SPS & & HP & SPS \\ \hline \\ [-8pt] 
     $a_{{\rm expt.}}$  & 2.906 & & 4.325  & 4.325       & & 4.243  & 4.242       \\ [+5pt]
     $c_{{\rm expt.}}$  & 2.837 & & \multicolumn{2}{c}{-}     & & \multicolumn{2}{c}{-}     \\ [+5pt]
     $a_{{\rm theo.}}$  & 2.922 & & \multicolumn{2}{c}{4.333} & & \multicolumn{2}{c}{4.247} \\ [+5pt]
     $c_{{\rm theo.}}$  & 2.847 & & \multicolumn{2}{c}{-}     & & \multicolumn{2}{c}{-}     
     \\ \hline \hline 
\end{tabular}
\end{center}
\label{lattice_parameter} 
\end{table}

%\clearpage 
\newcommand{\newblock}{}